# Orientation and structure of triple step staircase on vicinal Si(111) surfaces

S. A. Teys, K. N. Romanyuk, R. A. Zhachuk, and B. Z. Olshanetsky

*Institute of Semiconductor Physics, Siberian Division, Russian Academy of Science, Novosibirsk, 630090, Russia*

**Abstract**

The vicinal Si(111) surface, inclined towards the $[\bar{1}\,\bar{1}\,2]$ direction, was investigated by scanning tunnelling microscopy and spot profile analysing low energy electron diffraction. It has been established that the surface, consisting of regularly spaced triple steps and (111) terraces with a width equal to that of a single unit cell of the Si(111)-7×7 surface structure, has the (7 7 10) orientation. An atomic model of the triple step is proposed.

## 1. Introduction

In the process of self-assembling growth it is important to provide the formation of ordered and uniform nanostructures with high density and low dispersion in sizes. The use of high index surfaces and the stepped ones, in particular, as templates may facilitate the solution of this problem. Therefore, such surfaces and their possible application are intensively studied [1-9].

Vicinal Si(111) surfaces inclined towards the $[\bar{1}\,\bar{1}\,2]$ direction contain steps with the height of one and three interplanar distances $d_{111}$ (triple steps) at the temperatures below 870°C [10-12]. The share of triple steps increases with an increase of the inclination angle [13]. The most promising surface for the growth of high quality nanostructures is the surface which consists of only regular triple steps. According to Kirakosian *et al.* [14], the Si surface, which contains triple steps and regularly spaced (111) terraces with a width equal to that of a unit cell of the Si(111)-7×7 surface structure, has the (557) orientation. This conclusion is based on the analysis of high resolution scanning tunneling microscopy (STM) images. The Si(557) surface is inclined at 9.45° relative to the (111) plane towards the $[\bar{1}\,\bar{1}\,2]$ direction. The period of the staircase of triple steps on this surface is 5.73 nm [14]. However, in literature there is a discrepancy concerning the orientation of the triple step. Its orientation was determined as (112) [14] and as (113) [15]. The structure of the triple step was unknown until now.

The aim of present work was to determine the precise orientation of the triple step staircase containing regularly spaced $3d_{111}$-high steps and to explore the atomic structure of a triple step by low energy electron diffraction (LEED) and STM.

## 2. Experimental

The experiments were performed in two ultra high vacuum systems. One chamber was equipped with an STM (OMICRON) and the second one with a SPA-LEED (SPA is Spot Profile Analyzing) systems. STM images were recorded in the constant-current mode using an electrochemically etched tungsten tip. Silicon samples, inclined relative to the (111) plane towards the $[\bar{1}\,\bar{1}\,2]$ direction at 9.5° with an accuracy of ±0.5°, were used. Samples were resistively heated by a direct current. The current direction was parallel to the steps on the vicinal Si(111) surface to avoid the electromigration induced step bunching. The sample temperature was controlled by an optical disappearing-filament pyrometer. Clean surface was prepared following the procedure, described by Kirakosian *et al.* [14]. First a sample was carefully degassed at 600°C for several hours. Then, after short flash at 1250°C, the temperature of the sample was lowered to 1060°C in 30 s. After that the sample was quenched to 830°C, annealed at this temperature for 15 min, and cooled down to room temperature during 20 min.

## 3. Results

### 3.1. Orientation of the surface with regular triple steps

The surface consists of regularly spaced steps with the height of three interplanar distances (111) and the Si(111) terraces with the 7×7 surface structure. An atomic resolution STM image of an area of such surface is shown in Fig. 1. Sometimes in STM images one can see small surface areas with the metastable 5×5 surface structure and randomly distributed (111) terraces with the width equal to that of 2-3 unit cells of the 7×7 surface structure.

According to the STM data (Fig. 2), the projection of a period of the triple step staircase on the (111) plane equals to 16 distances between the <110> rows. So this projection corresponds to the $[\bar{4}\,\bar{4}\,8]$ vector in the cubic lattice. The period of the staircase equals to the difference between the [$\overline{4}48$] and [111] vectors. Hence, it corresponds to the [$\overline{5}\overline{5}7$] vector, and the orientation of the surface with a staircase is (7 7 10).

To check the staircase orientation, derived from STM images, we used SPA-LEED technique [16]. Fig. 3a shows a LEED pattern from the sample surface obtained at 61 eV. In Fig. 3b the positions of the diffraction spots along the [$\overline{1}\,\overline{1}\,2$] miscut direction through the position of (00) spot from the (111) terraces are shown versus the scattering vector $k_\perp$ and the energy of electrons. In our system we could not get diffraction patterns at the electron energies greater than 244 eV. To make the image more comprehensible, we extended the upper part of Fig. 3b. Figure 3b represents a vertical cut of reciprocal lattice of the vicinal Si(111) surface with $k_\parallel$ as the $x$ axis along the [$\overline{1}\,\overline{1}\,2$] direction and $k_\perp$ as the $y$ axis in the [111] direction. The tilted rods are produced by movement of the diffraction spots from the vicinal surface at the change of the electron energy. These rods are normal to the staircase surface. The surface orientation can be determined by calculating the difference between the indices of the positions of the vertical rods of ($0\,\overline{1}$), (00), and (01) spots from the (111) terraces, at which they are crossed by an inclined rod. In our case a tilted rod crosses the position with the indices (8 8 10) - (111)/3 = (23 23 29)/3. Therefore, the surface orientation is (23 23 29)/3 - (333), which is equivalent to (7 7 10). This plane makes an angle of 10.02$^\circ$ with the (111) plane. Thus, the results, obtained by LEED, confirm that the surface with regular triple steps has (7 7 10) orientation, and not the (557) one, which was supposed earlier [6, 14].

### 3.2. Atomic structure of a triple step

To reveal the atomic structure of the triple steps we used STM images obtained at different polarities of the applied bias (Figs 2, 4). A tentative model of a triple step was developed comparing possible configurations of Si atoms with the high resolution STM images. The model is presented in Fig. 5a. The details of the model are described below.

The (111) terraces in the staircase contain an additional row of $R$ atoms at the bottom of a triple step (Fig. 2) [17, 18]. Therefore the width of a terrace is equal to that of a single unit cell of the 7×7 surface structure plus two <110> interrow distances. On the edge of the upper (111) terrace there are $A_3$ atoms and $D^{\parallel}_3$ parallel dimers. Here lower indices indicate a number of the bilayer in a triple step, at which the corresponding element is located. Upper indices indicate orientation of the dimers relative to the step edge. The bilayer numeration starts from the bottom of a step. The $A_3$ atoms are the adatoms in the Si(111)-7×7 unit cell [19]. They have three saturated and one dangling bonds.

At the level of the second bilayer of a triple step there is a row of the $A_2$ adatoms with the periodicity of $2 \times I_{110}$ ($I_{110} = 0.383$ nm). The diffraction spots from the $2 \times I_{110}$ periodicity can be seen on the diffraction pattern (Fig. 3a). Similar spots were observed earlier [15], but were left unexplained. The $A_2$ atoms also have three saturated and one dangling bonds. The $A_2$ atoms are located either against the $A_3$ atoms or against the gaps between the $A_3$ atoms, because the periodicity of the $A_3$ atoms in the $[1\bar{1}0]$ direction is $7 \times I_{110}$, whereas the periodicity of the $A_2$ atoms is $2 \times I_{110}$. According to STM data, the distance between the $A_3$ and $A_2$ atoms (Fig. 2) is $1.04 \pm 0.03$ nm. This is approximately 0.15 nm bigger than the corresponding distance in the ideal crystal lattice of Si. This discrepancy may be due to the distortion of the crystal lattice at the edge of a triple step or to the redistribution of electronic density in $A_2$ atoms.

STM images often reveal a row defect ($RD$) in $A_2$ chains (Figs 1, 2, 4). The scheme of a row defect is shown in Fig. 5b. Its configuration is as follows. One of the $A_2$ atoms shifts by one row in the direction of upper terrace and by half of a period in the $[1\bar{1}0]$ direction (Fig. 5b). In the area of an $RD$ the distance between neighboring atoms in the $A_2$ chain becomes $3 \times I_{110}$. The $RD$ develops only against the corner vacancies of the $7 \times 7$ surface structure on the upper terrace. In Fig. 1 the row defects can be seen against approximately 25% of the vacancies. The symmetric positions of the $A_2$ atoms relative to $RD$ defects imply that the lower atoms at the second bilayer are not shifted from their normal positions in the ideal crystal lattice of Si.

At the negative bias polarity the spot corresponding to the dangling bond of a rest atom, close to the $RD$ defect, becomes brighter (Fig. 4b). The images of other rest atoms in the second bilayer of the triple step are also visible. These rest atoms are located between the $D^{\parallel}_3$ dimers of the $7 \times 7$ surface structure and the chain of the $A_2$ atoms (Fig. 5b). The images of the rest atoms in the second bilayer, including those in the area of the $RD$ defect, have different brightness, owing to different configuration of their closest neighbors.

Below the $A_2$ adatoms the rows of perpendicular ($D^{\perp}$) and of parallel ($D^{\parallel}_1$) dimers are located (Fig. 5a). The periodicity of the $D^{\perp}$ dimers is $1 \times I_{110}$, and that of the $D^{\parallel}_1$ dimers is $2 \times I_{110}$. An assumption that the $D^{\perp}$ dimers form on the edge of a triple step on the vicinal Si(111) surface inclined towards $[\bar{1}\bar{1}2]$ was made earlier [20, 21]. The possibility of the formation of $D^{\perp}$ and $D^{\parallel}$ dimers along the steps on the Si(100) surface was considered by Chadi [22]. The $D^{\perp}$ dimers on the triple step are not symmetric. The upper atom of a $D^{\perp}$ dimer has one dangling bond, whereas all bonds of the lower atom are saturated (Fig. 5a).

The brightness distribution in the image of the $D^{\perp}$ dimer chain depends on the bias polarity (Figs 4a and b). At the positive bias polarity (tunneling into the empty states of the surface) the images of the dimers, located just against the gaps between $A_2$ adatoms, are the brightest. At the negative polarity (tunneling from the filled states of the surface) the brightness distribution in the images changes, and the images of the dimers, located against the $A_2$ atom positions, become the brightest (Fig. 4b). The shape and the brightness of the $D^{\perp}$ dimer images are affected by the neighbouring $A_2$ atoms. In Fig. 4b it is seen that they depend on whether a $D^{\perp}$ dimer is located near an $RD$-defect or not.

The $[1\bar{1}0]$ row of $D^{\parallel}_1$ dimers is located at the level of the first bilayer of a triple step, below the $D^{\perp}$ dimer row (Figs 2, 5a). The $D^{\parallel}_1$ dimers are visible at both polarities of applied bias (Figs 4). At the positive bias polarity (Fig. 4a) we can see the dangling bonds of the dimer atoms as a couple of bright maxima. At the negative bias polarity (Fig. 4b) the saturated covalent bond between the two dimer atoms is visible. The dependence of brightness versus bias polarity is the most apparent in the right part of Figs 4a and b, near the dark area. The dark area is the image of vacancy defect, which is the absence of two $D^{\perp}$ dimers. Similar changes in the STM images of the dimers on the Si(100) surface as a function of polarity of applied bias were observed earlier [23-25].

The careful analysis of the STM images and of the possible positions of Si atoms reveals the zigzag-like row of atoms with a $1 \times I_{110}$ periodicity. This row is marked as $ZR$ in the model in Fig. 5a. The zigzag row is located on the lower terrace at the bottom of a triple step and at the level of the R atoms. Every atom in the $ZR$ has one dangling bond, one bond connected with the atoms of lower layer, and two bonds connected with neighboring atoms in the row (Fig. 5a). Earlier such a row was observed along regular triple steps as well as along single triple steps on clean vicinal Si(111) surfaces inclined towards the $[\bar{1}\bar{1}2]$ direction at different angles [21]. At both polarities of applied bias the brightness of $ZR$ atoms in the STM images is close to that of $D^{\parallel}_1$ dimers and of $R$ atoms at the bottom terrace. Hence, the $ZR$ atoms are located at the level of the $D^{\parallel}_1$ dimers with an accuracy of 0.05 nm. At the same time they are 0.2-0.3 nm higher than the rest atoms of the lower terrace. Change of the positive bias polarity to the negative one causes the shift of the positions of $ZR$ image maxima in the $[1\bar{1}0]$ direction by half of the $I_{110}$ period (Fig. 4). The new positions of the brightness maxima are located under those of the $D^{\perp}$ dimers. The positions of the bright maxima relative to the bright spots of the $R$ atoms of the lower terrace also change respectively. One can see, that at the negative bias the STM image of $ZR$ is shifted in the $[\bar{1}\bar{1}2]$ by 0.1 nm (Fig. 4b) relative to the position of the image at the positive sample bias. Shift of the bright maxima at the change of the bias polarity can be attributed to the asymmetric distribution of electronic density in the *ZR*.

Taking into account the revealed atomic structure, no definite crystallographic plane can be assigned to the triple step.

## 4. Conclusions

The vicinal Si(111) surface, inclined relative to the (111) plane towards the $\left[\bar{1}\,\bar{1}2\right]$ direction, was investigated by STM with atomic resolution and SPA-LEED. It was established that the surface, consisting of regularly spaced triple-layer-height-steps and (111) terraces with a width equal to that of a single unit cell of the Si(111)-7×7 surface structure, has the (7 7 10) orientation. A tentative atomic model of the triple step is proposed on the basis of high resolution STM images. The model includes parallel and perpendicular dimers, an additional *R* row of atoms and a zigzag-like row of atoms at the bottom of a triple step.

## Acknowledgements

This work was supported by the Russian Foundation for Basic Research and by the Ministry of Education and Science of Russia (Russian Federal Program).

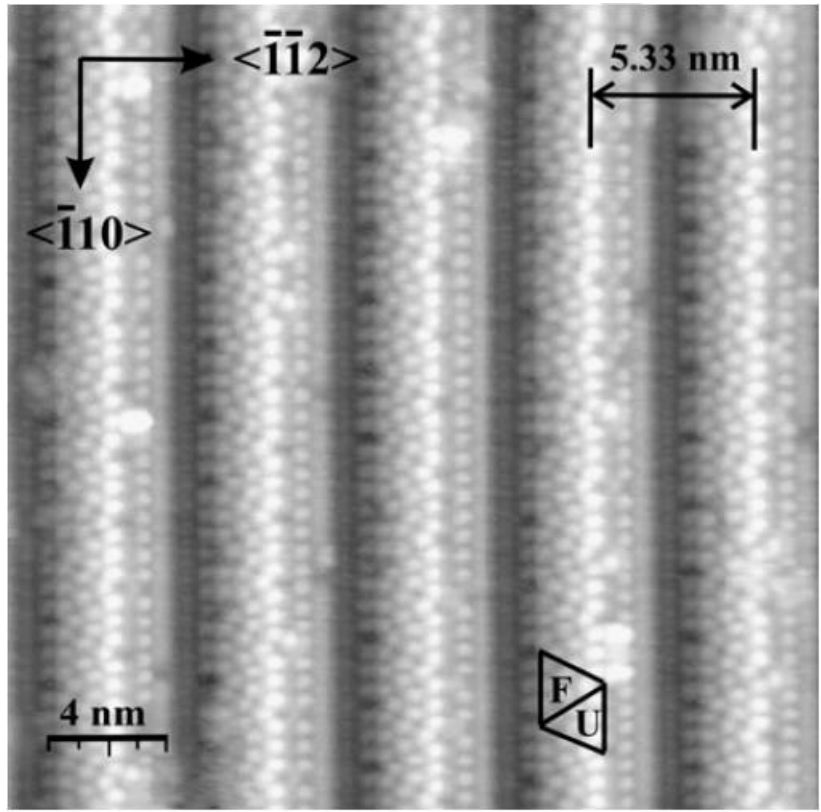

Fig. 1. STM image of a vicinal Si(111) surface consisting of (111) terraces with a width of that of the Si(111)-7×7 unit cell. The terraces are separated by 3 $d_{111}$-high steps. Sample bias U = + 2.0 V (unoccupied surface states). F is faulted half of cell; U is unfaulted half of cell.

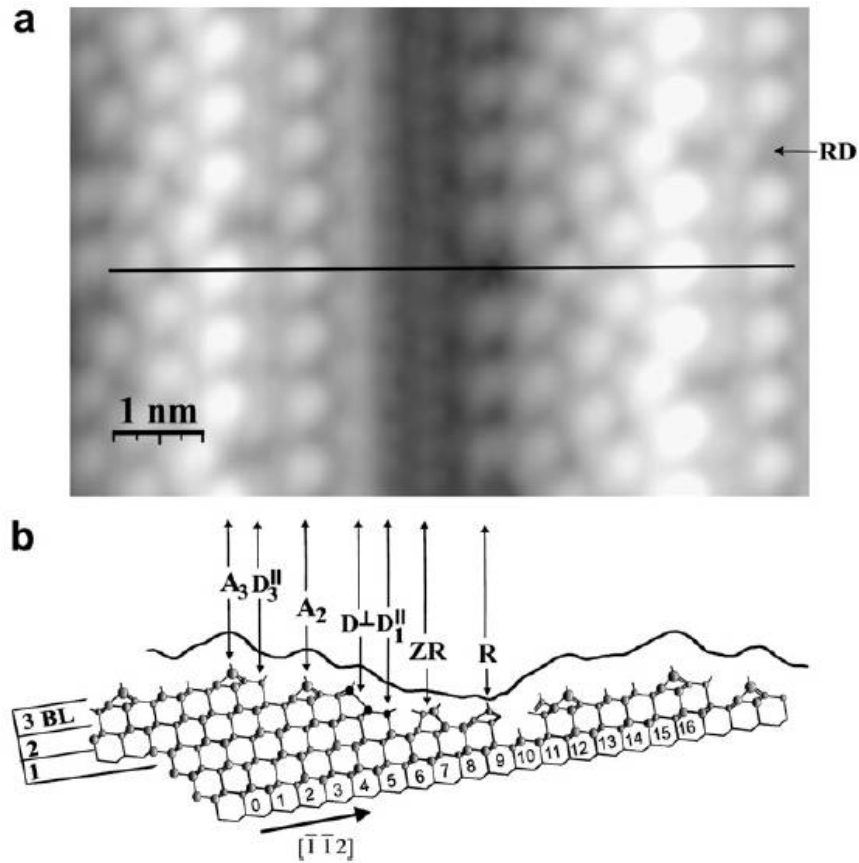

Fig. 2. Triple step. (a) STM image of a triple step. Sample bias U = + 1.0 V (unoccupied surface states). (b) Profile of the surface (wavy line) and a scheme of the crystal lattice cross-section along the black line in Fig. 2a.

$A_3$ is an adatom of the (111)-7×7 structure; $D^{\parallel}_3$ is a dimer parallel to the step edge; $A_2$ is an atom in the $<1\bar{1}0>$ row at a level of the second bilayer; $D^{\perp}$ is a dimer perpendicular to the step edge direction; $D^{\parallel}_1$ is a dimer parallel to the step edge at the level of the first bilayer; $ZR$ is a zigzag like row of atoms and $R$ is an additional adatom row at the bottom of a step; $RD$ is a row defect.

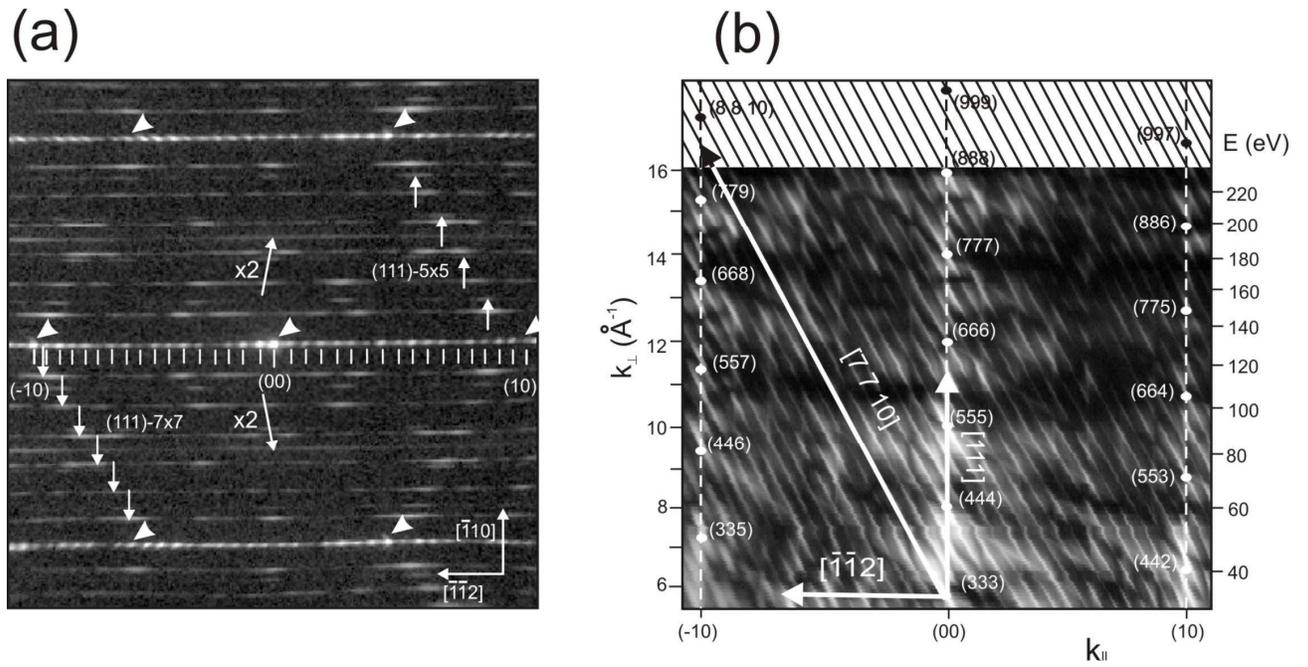

Fig. 3. LEED patterns from Si surface. (a) Diffraction spots from the Si(111)-7×7 and Si(111)-5×5 structures and those from the rows of $A_2$ atoms are indicated by arrows. Integral order spot positions are marked by triangles. Electron energy is 61 eV. (b) Vertical cut of reciprocal lattice of the vicinal Si(111) surface in the $<\bar{1}\,\bar{1}\,2>$ direction.

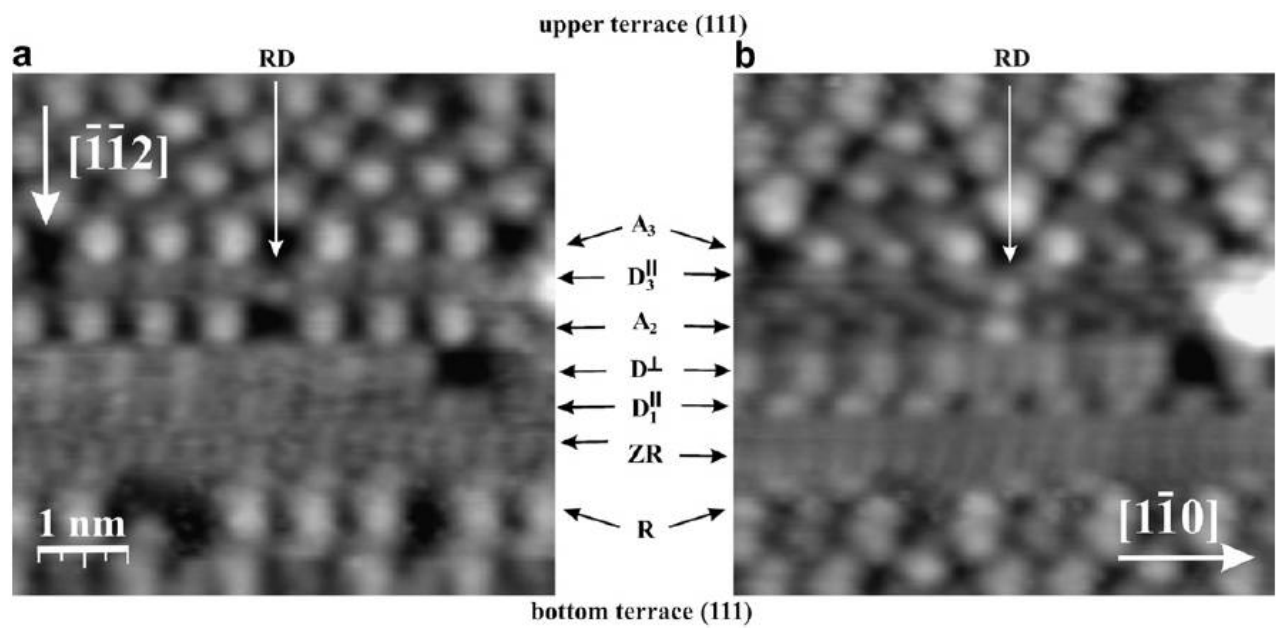

Fig. 4. STM images of an area of a triple step at sample bias: (a) U = +1.0 V (unoccupied surface states); (b) U = - 1.4 V (occupied surface states).

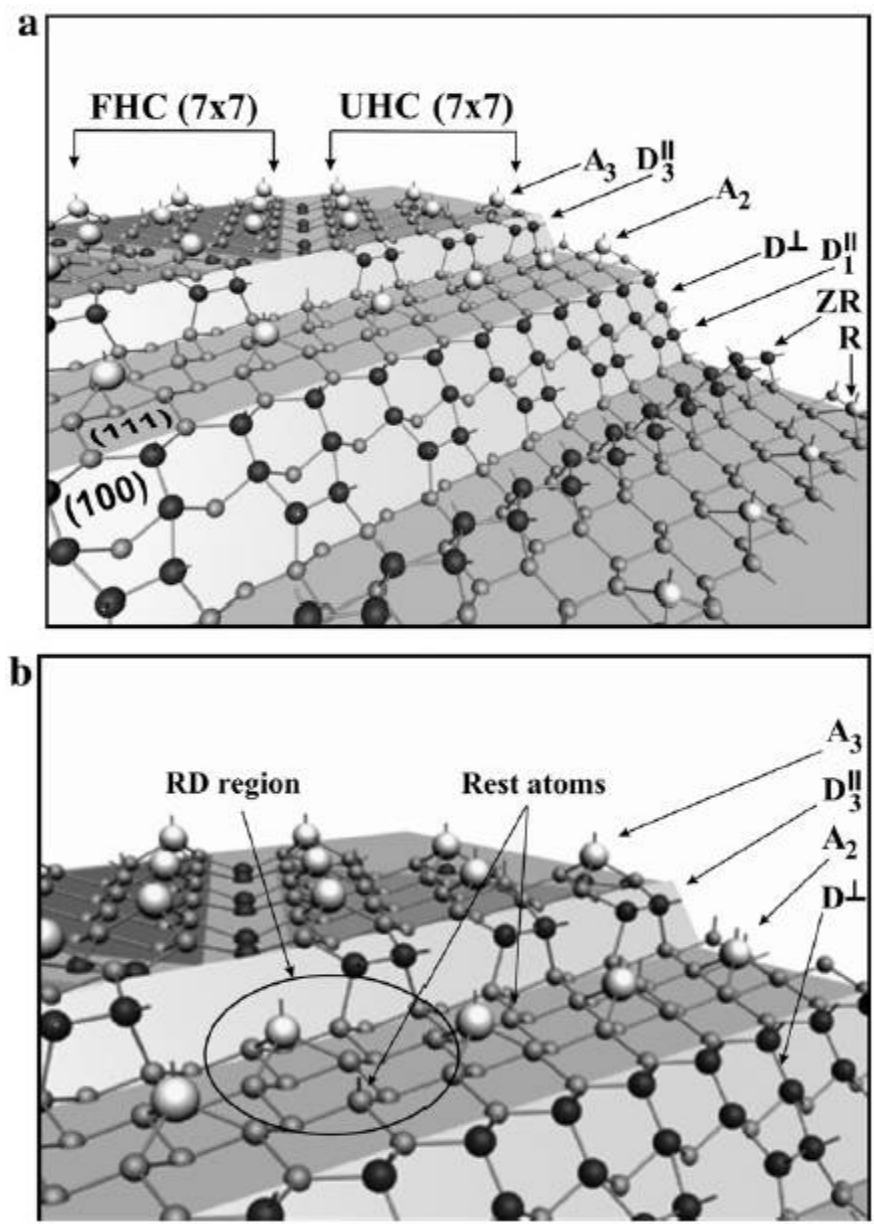

Fig. 5. Models: (a) Triple step. FHC is faulted half of cell; UHC is unfaulted half of cell.
(b) Row defect (*RD*).